\def\ube13{UBe$\rm_{13}$}

\def\bi2212{Bi$\rm_2$Sr$\rm_2$CaCu$\rm_2$O$\rm_8$}
\def\ybi2212{Bi$\rm_2$Sr$\rm_2$YCu$\rm_2$O$\rm_8$}
\def\ycabi2212{Bi$\rm_2$Sr$\rm_2$Ca$\rm_{1-x}$Y$\rm_x$Cu$\rm_2$O$\rm_{8+\delta}$}
\def\ybco{YBa$\rm_2$Cu$\rm_3$O$\rm_{6.9}$}
\def\y65cabi2212{Bi$\rm_2$Sr$\rm_2$Ca$\rm_{0.35}$Y$\rm_{0.65}$Cu$\rm_2$O$\rm_{8+\delta}$}

\documentstyle[aps,prb,twocolumn,epsf]{revtex}
\begin{document} 
\draft

\def\dfrac#1#2{{\displaystyle{#1\over#2}}}
\twocolumn[\hsize\textwidth\columnwidth\hsize\csname @twocolumnfalse\endcsname
Submitted to Phys. Rev. B Rapid Commun. \hfill{LA-UR-00-2381}

\title{Low Temperature Thermal Transport in High Temperature Superconducting Bi$_2$Sr$_2$CaCu$_2$O$_8$ via Y-doped Insulating Analogue}

\author{R. Movshovich,$^1$ E. G. Moshopoulou$^*$,$^{1,2}$ P. Lin,$^3$ M. B. Salamon,$^3$ M. Jaime,$^1$ M. F. Hundley,$^1$ and J. L. Sarrao$^1$ }
\address{$^1$Los Alamos National Laboratory, Los Alamos, New Mexico 87545 \\ $^2$Brookhaven National Laboratory, Upton, NY 11973 \\ $^3$University of Illinois at Urbana-Champaign, Urbana, Illinois 61801
}

\date{\today}

\maketitle

\begin{abstract} 

In order to separate electronic and phonon mechanisms of thermal transport in high temperature superconducting \bi2212, we measured thermal conductivity of single crystals of Y-doped \bi2212, an insulating analogue of pure \bi2212, in the temperature range between 40 mK and 2 K. Temperature dependence of the thermal conductivity of insulating samples changes from T$^{1.7}$ between 1 and 2 K to $T^{2.6}$ below 200 mK, indicating saturation of the phonon mean free path with decreasing temperature. Overall, thermal conductivity of the insulating samples (which is due entirely to phonons) is strongly suppressed in comparison to that of the superconducting samples. Therefore, a large (and at the lowest temperatures, predominant) part of the total thermal transport in superconducting \bi2212 is due to electrons. 

\end{abstract}


]
\narrowtext

Thermal conductivity has become a popular tool to study the symmetry of the order parameter in unconventional superconductors. The reason for this lies in the particular sensitivity of thermal transport to the normal quasiparticle spectrum deep in the superconducting regime, because the superconducting condensate itself does not carry heat. Theoretical investigations predicted universal (impurity concentration independent) electrical and linear-in-temperature contribution to the thermal  conductivity for a superconductor with a $d_{x^2-y^2}$ order parameter.~\cite{lee:prl_93,herschfeld86,schmitt86,graf96:tc} Such behavior is understood as a result of the cancelation between increases in both the bandwidth of electronic impurity states and in impurity scattering of quasiparticles along the nodal directions of a $d_{x^2-y^2}$ superconducting order parameter. Detailed investigation of the low temperature thermal conductivity of Zn-doped YBa$_2$Cu$_3$O$_{6.9}$ indeed revealed such a behavior,~\cite{taillefer:prl_97} with the absolute value of electronic thermal conductivity within a factor of two of expected value, a validation of the developed theories of impurity scattering in the High Temperature Superconductors (HTS).

Low temperature thermal conductivity was also measured in another HTS compound, \bi2212, with some unexpected results. Magnetothermal conductivity was first measured for temperature down to 6 K and field up to 14 T.~\cite{krishana97} After the initial decrease of thermal conductivity $\Delta \kappa$ of a few percent, a sharp kink in thermal conductivity was observed at a critical field of $H_k$, with thermal conductivity becoming field-independent above $H_k$. Such behavior was observed for the temperature range between 6 and 20 K, with both $\Delta \kappa$ and $H_k$ proportional to $T^2$. These results were interpreted as a phase transition in the condensate, with all of the Fermi surface gapped in the new high field state. Within this picture the quasiparticles (electrons) were therefore responsible only for a few percent of the total thermal conductivity $\Delta \kappa$ at zero field. Several groups were able to both reproduce and extend measurements of magnetothermal transport in \bi2212.~\cite{aubin:prl_99,ando:preprint_99} Observation of hysteresis in magnetic field, sample-dependent behavior, and simultaneous theoretical work~\cite{franz:prl_99,vekhter:prl_99} lead to the conclusion~\cite{aubin:prl_99,ando:preprint_99} that it is the vortex scattering of normal quasiparticles that is responsible for the apparent field plateau feature for temperature above few K. Recent theoretical work by Vekhter {\it et al.}~\cite{vekhter:prl_99} concludes that the ``plateau" is the result of the transition from the temperature dominated regime (at low field) to the field dominated regime above few Tesla, where the electronic thermal conductivity varies as ${\sqrt{H}}ln(H_{c2}/H)$. 

The low temperature thermal conductivity (below few hundreds of mK) was shown to follow the $\sqrt H$ behavior both in \ybco~\cite{chiao:prl_99} and \bi2212.~\cite{aubin:prl_99} However, the much larger magnitude of the $\sqrt{H}$ term in the latter compound was puzzling,~\cite{chiao:prl_99} since the \ybco \ single crystals are generally considered to be of higher quality on the basis of the normal state properties, such as residual resistivity.

In an unrelated experiment, a sharp drop of thermal conductivity in two Ni-doped \bi2212 samples was observed in zero field at about 200 mK.~\cite{movshovich:prl_98} It was suggested that this feature is a consequence of a bulk phase transition into a second superconducting $d_{x^2-y^2}+id_{xy}$ (so called $d+id$) state,~\cite{movshovich:prl_98,balatsky:prl_98} where the nodes of the $d_{x^2-y^2}$ state are lifted, and the energy gap is finite around the whole Fermi surface. For this interpretation to be valid, it is necessary that most of the heat current in \bi2212 at low temperature (below few degrees K) is carried by electrons. 

Interpretations of the experimental results of a number of thermal transport studies of \bi2212 described above rely crucially on the ratio of electronic to phonon heat currents. It would therefore be of great value to 

\begin{figure}
\epsfxsize=3in
\centerline{\epsfbox{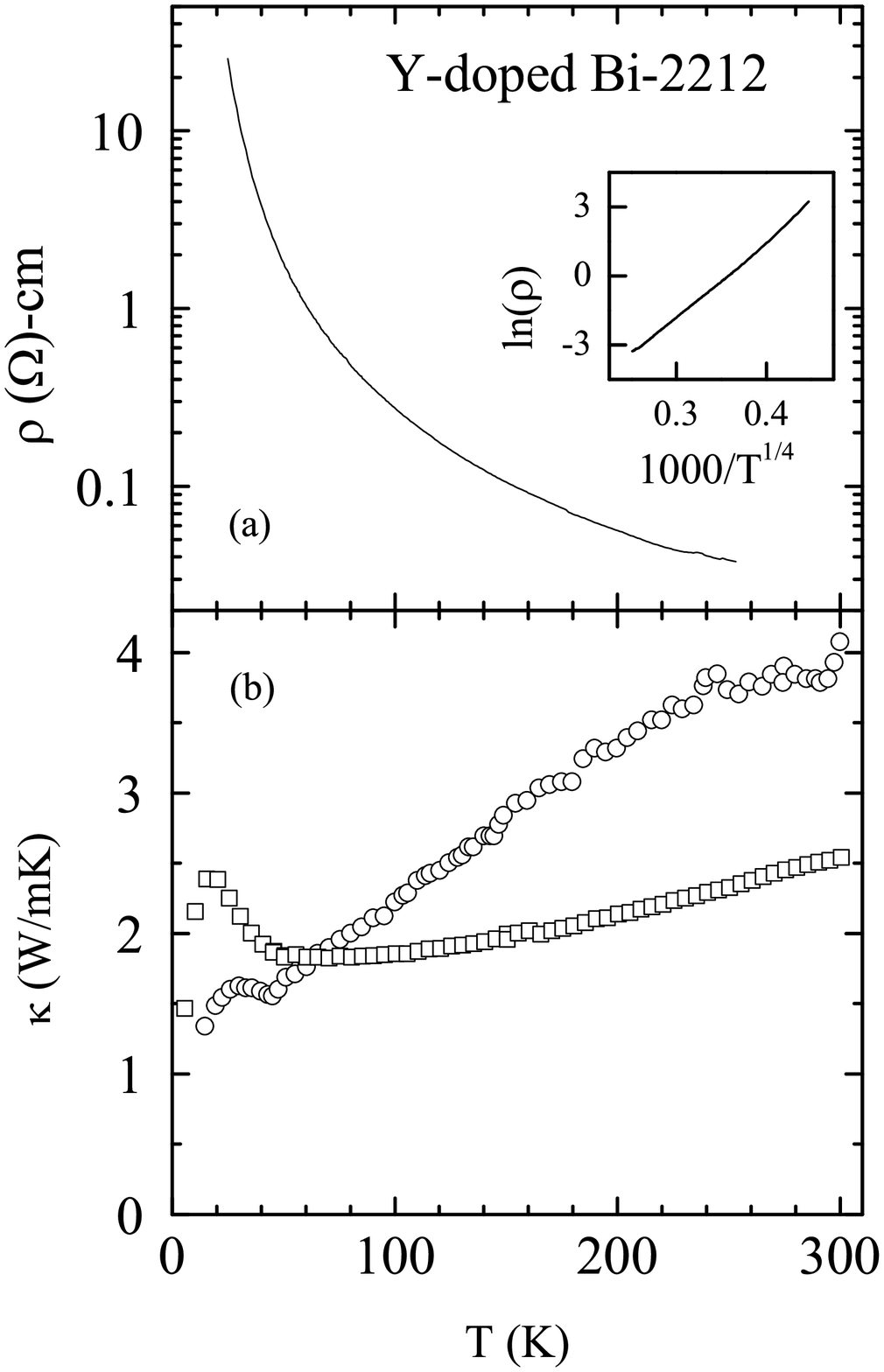}}
\caption{Room temperature to 5 K transport in \y65cabi2212. (a) Resistance of \y65cabi2212 sample vs. temperature, showing insulating behavior with $dR/dT < 0$. (b) ($\circ$) thermal conductivity of \y65cabi2212 sample used in the very low temperature studies (see below); ($\Box$) thermal conductivity of Y-doped insulating \bi2212 from ref. 14.}
\label{resistance}
\end{figure}

\noindent separate these contributions experimentally.

The background phonon thermal conductivity in the superconducting \ybco was measured directly by deoxygenating a pure superconducting sample, thereby driving it insulating.~\cite{taillefer:prl_97} The thermal conductivity of this insulating analogue was taken as a measure of the phonon thermal conduction in the superconducting sample. It was determined in this way that the phonons carried a major part of the heat current above about 100 mK, and that all of the increase of the thermal conductivity above the universal value for temperature up to a few hundred millikelvin can be accounted for by phonons.

We took another approach to determine the phonon thermal conductivity in \bi2212, substituting Y$^{3+}$ for Ca$^{2+}$ (and thereby removing holes from the conduction band) to drive it across the superconductor/insulator transition. Such approach was used previously to study the phonon thermal conductivity in both \bi2212 and \ybi2212 compounds between 10 and 300 K.~\cite{allen:prb_94} This study demonstrated that the thermal conductivity in the insulating \ybi2212 represents very closely the phonon background thermal conductivity of \bi2212 in the normal state. As the temperature is lowered below T$_c$, the influence of crystalline point imperfections in the doped samples should decrease even further, since the long wavelength phonons, insensitive to point defects, become the dominant heat carriers. Therefore we expect the thermal conductivity of the insulating Y-doped \bi2212 samples to be a very good representation of the phonon thermal conductivity of the superconducting \bi2212. 

The insulating crystals used in our study were \ycabi2212 with $x \approx 0.65$, determined as described below. This composition was chosen on the basis of the following considerations. Higher Y-doping lowers the quality of the crystals. Indeed, several groups~\cite{mitzi:prb_90,kendziora:prb_92,soto:prb_96,manifacier:physicaB_99} reported that highly Y-doped crystals present``islands" of distinctly different dopant concentration, vacancies, and twinning. Therefore, our strategy was to substitute just enough Y$^{3+}$ for Ca$^{2+}$ in order to suppress superconductivity, and keep the highest crystalline homogeneity and quality of the samples. Single crystals of \ycabi2212 were grown by the self-flux technique, using the directional solidification approach described by Mitzi et al.~\cite{mitzi:prb_90} It is known for this system that actual (measured) crystal composition is distinctly different from the starting (nominal) composition. The starting materials were taken in atomic ratio of 2.4Bi:2Sr:0.5Ca:0.5Y:2.4Cu. Bi$_2$O$_3$ and CuO were included in excess to act as flux for the crystal growth, and to compensate for the evaporation of Bi$_2$O$_3$ during the growth. The resulting mica-like crystals were easily removed from the melt. 

We characterized the sample by performing synchrotron X-ray powder diffraction measurement on the X7A beamline at the National Synchrotron Light Source at Brookhaven National Laboratory. The diffraction peaks' shape and the evolution of their full-width at half-maximum (FWHM) as a function of q give us insight into the homogeneity of the samples, whereas determination of the lattice constants give information on the average Y-content of the crystals.
The X-ray diffraction spectrum was compared with the theoretical one calculated using the software LAZY-PULVERIX40. This comparison revealed that most observed diffraction lines correspond to \ycabi2212. Some extra weak peaks observed in the experimental pattern are expected, and are due to flux or secondary material remaining on the surface of some crystals.~\cite{moshopoulou:tobepublished} The results obtained from these fits are the following:
Firstly, the diffraction peaks do not have a systematically asymmetric shape, which would suggest slight inhomogeneities in composition. Furthermore, there are no weaker peaks closely adjacent to, but distinguishable from, the main peaks (which would reflect compositional segregation). It is worth noting here that Mitzi {\it et al.}~\cite{mitzi:prb_90} observed such a secondary group of diffraction lines for highly Y-doped crystals ($x \approx  0.8$ to 1), even with the laboratory x-rays which have one order of magnitude lower resolution and peak-to-background

\begin{figure}
\epsfxsize=3in
\centerline{\epsfbox{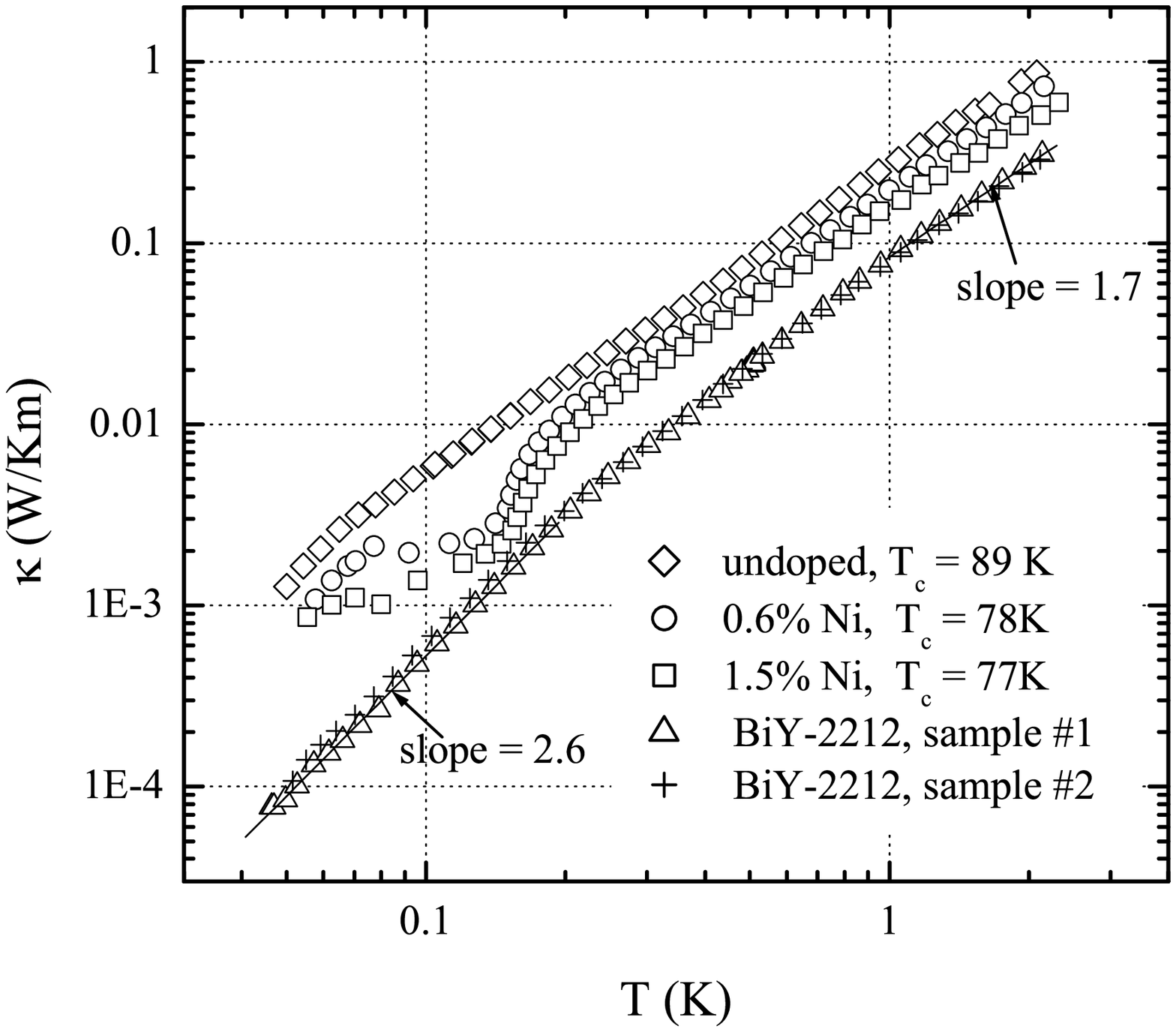}}
\caption{Thermal conductivity vs. temperature of \bi2212 samples. Superconducting samples: ($\Diamond$) undoped; ($\circ$) 0.6\%Ni-doped; ($\Box$) 1.5\%Ni-doped. Insulating samples: (+,$\triangle$) \y65cabi2212 \#1 and \#2.}
\label{tc}
\end{figure}

\noindent ratio than the synchrotron x-rays used in this work. Secondly, by expressing the evolution of the FWHM of the peaks as a function of $2 \Theta$, and using the Williamson-Hall analysis,~\cite{williamson:53}  a $\Delta c/c$ of $7.3 \times 10^{-4}$ was deduced, which corresponds to a compositional inhomogeneity $\Delta x < 2.4$\%. This result is consistent with the sharpness of the diffraction peaks of the sample (the FWHM of most peaks is about 3 to 4 times that of Si), an observation which also demonstrates the high quality of the sample. Finally, the cell constants were obtained by a least-squares refinement of the positions $2 \Theta$ of 15 reflections. The resulting parameters are a = 5.453(3) \AA, b = 5.409(4) \AA, and c = 30.302(11) \AA, which result in $ x \approx 0.65$ according to the well established correlation  between the lattice constants and Y concentration x.~\cite{kendziora:prb_92}

Figure~\ref{resistance}(a) shows low temperature resistivity of one \ycabi2212 sample with $x \approx 0.65$ versus temperature. This sample is clearly insulating, with the negative temperature derivative of resistance in the whole temperature range from room temperature down to 4 K. The inset shows that $\ln R$ is close to linear in $1/T^{1/4}$, which is characteristic of the Variable Range Hopping (VRH) conduction mechanism. Substituting Y$^{3+}$ for Ca$^{2+}$ in \bi2212 not only removes holes from the conduction band but also, because it takes place in the immediate neighborhood of the Cu-O planes, generates a random potential felt by the charge carriers. It is therefore not surprising that resistivity follows the VRH dependence. Similar conclusions were reached in several previous studies,\cite{yoshizaki:physicaC_88,fukushima:jjap_88,tamegai:jjap_89,mandal:prb_91}, even though the question of the dimensionality (2D or 3D) of the VRH conduction remains under debate.

Figure~\ref{resistance}(b) shows the thermal conductivity between 300 K and 5 K of the \y65cabi2212 sample used for the ultra low temperature measurements (see below). For comparison, we also plotted the digitized thermal conductivity data of insulating Y-doped \bi2212 sample from the previous work by Allen {\it et al.}~\cite{allen:prb_94} Thermal conductivity of both samples display similar trends, with monotonic decrease upon cooling from room temperature followed by a peak around 10 to 15 K. Samples appear to be of comparable quality, with thermal conductivities within a factor of two in the whole temperature range investigated. However, the low temperature end of the data for the sample used in the present study is about 80\% greater, and, therefore, represents more accurately the low temperature phonon thermal conductivity in pure \bi2212.

Figure~\ref{tc} shows the thermal conductivity data for two insulating samples from the same growth as the resistance sample discussed above. Also for comparison we show the data collected and discussed previously on three superconducting samples, grown by the Traveling Solvent Floating Zone method.~\cite{movshovich:prl_98} The data for the two insulating samples overlap rather well, even though the geometrical factors for these samples are different by a factor of two. The temperature dependence changes from $T^{1.7}$ above 1 K to $T^{2.6}$ below $\approx 250$ mK, indicating increase of the phonon mean free path with decreasing temperature. Phonon thermal conductivity with temperature-independent mean free path should be proportional to $T^3$, and this regime is evidently approached below 250 mK. 

The~mean~free~path~in~the~case~of \y65cabi2212 is not, however, determined by scattering off the sample boundaries, in contrast to the case of \ybco. There it was shown that all of the low temperature rise of the thermal conductivity (below 200 mK) can be accounted for by the sample-size-limited phonon thermal conductivity.~\cite{taillefer:prl_97} Such an estimate for the insulating samples shown in Fig.~\ref{tc} falls more than an order of magnitude above the experimental data. Therefore, the mechanism which leads to the saturation of the phonon thermal conductivity at low temperatures is not the surface boundary scattering. The high quality of the single crystals of \ycabi2212, as demonstrated by the synchrotron X-ray diffraction analysis, makes phase segregation an unlikely reason for the increase in phonon scattering. Substantially shortened mean free path must be an intrinsic property of \bi2212 and related compounds. Strong phonon scattering in insulating \ybi2212 has been noted previously,\cite{allen:prb_94} where the authors conclude that the phonons are either poorly defined or non-existent at room temperature. The overall temperature behavior of thermal conductivity in \ybi2212 (and by extension, \bi2212, its superconducting analogue) resembled more that of glasses than of good crystalline insulators.

Thermal~conductivity~of~insulating \break \y65cabi2212 samples is about an order of magnitude smaller than that of the superconducting \bi2212 samples at the lowest temperature, and is below it in the whole temperature range studied (see Fig.~\ref{tc}). Therefore, the majority of the heat current in superconducting \bi2212 must be carried by the electrons. 
A consistency check can be made by estimating the electronic mean free path (mfp) that will account for the temperature dependence of the low temperature thermal conductivity in superconducting \bi2212. We use the expression for thermal conductivity of the $d$-wave superconductor in 2D arrived at by Krishana {\it et al.}~\cite{krishana97},

\begin{equation}
\kappa^{2D} = \eta{{k_B^2 T}\over {\hbar}}{{k_B T}\over {\Delta}} k_F l
\end {equation}

\noindent where $k_F$ is a Fermi wave vector and $l$ is a quasiparticle mean free path (mfp), to estimate the electronic mpf and scattering rate for pure \bi2212. Recent results on microwave conductivity on \bi2212~\cite{turner:cond_mat} validate the assumption of the energy independence of the quasiparticle relaxation rate relied upon in derivation of eq. (1). We use the data in Fig. 2 of the ref.~\onlinecite{movshovich:prl_98}, where $\kappa / T$ is plotted versus T to highlight the low temperature $T^2$ temperature dependence, and obtain $l \approx 6$ $\rm\mu m$. We can compare this  value with experimental values for the HTS available in the literature. Mean free path for several optimally doped \ybco \ samples was determined via thermal Hall effect measurements at temperatures above 10 K.~\cite{krishana:prl_95,krishana:prl_99} Longitudinal mean free path $l$ grows dramatically below the superconducting transition temperature $T_c$, from 550 to 4000 \AA \ between 50 and 10 K. The overall dependence in this temperature range is close to $l \propto T^{-1}$. The mfp does not show any sign of saturation in the temperature range studied. The temperature range of the thermal conductivity data used for our estimate of the low temperature quasiparticle mfp is two orders of magnitude lower (between 50 and 250 mK) than in ref.~\onlinecite{krishana:prl_95,krishana:prl_99}. Therefore, mfp of about one order of magnitude larger, in the range of microns, seems to be reasonable and consistent with the observed trends at higher temperatures. More recent microwave spectroscopy measurements of pure \ybco samples determined the quasiparticle mfp to be 4 $\mu$m,~\cite{hosseini:prb_99} of the same order as the estimate we obtain for \bi2212.

The discussion above of the estimated quasiparticle mfp supports the conclusion we drew from the thermal conductivity measurements of the insulating analogue \ybi2212: that the low temperature thermal transport in pure \bi2212 at low temperature is dominated by quasiparticles. One of the consequences of this is that the 200 mK anomaly in thermal conductivity of the two Ni-doped \bi2212 samples, observed previously, must be of electronic origin, in support of the $d + id$ hypothesis of the origin of this anomaly.~\cite{movshovich:prl_98} 

The predominance of the electronic thermal conductivity at low temperature in \bi2212 may help to interpret several other recent experimental findings as well. For example, a larger relative increase in the low temperature thermal conductivity in \bi2212 in comparison with \ybco (as discussed above) is likely due to the much larger role the quasiparticle thermal transport plays in \bi2212.

In conclusion, thermal conductivity measurements of the \ybi2212, an insulating analogue of the high temperature superconducting \bi2212, show that the low thermal transport in \bi2212 below $\approx 2$ K is dominated by electrons. This finding has important implications for our understanding of the low temperature properties of HTS, including zero field and magnetothermal transport. In particular, it contributes to a self-consistent picture of the very low temperature properties of HTS deep in the superconducting regime. In contrast, the difference between the normal state properties of \ybco and \bi2212, e.g. residual resistivity,  points to a lacking of unified picture of transport in HTS above $T_c$.

We thank I. Vekhter and A. V. Balatsky for many useful discussions. One of us (E. G. M.) thanks Zachary Fisk for his advice on the crystal growth and D. E. Cox for stimulating discussions on the synchrotron x-ray powder diffraction study. Part of this work was done at the Aspen Center for Physics, and one of us (R. M.) thanks the Center for the offered hospitality. Work at Los Alamos National Laboratory was performed under the auspices of the U.S. Department of Energy. Work at Brookhaven National Laboratory was supported by the U.S. Department of Energy, Division of Materials Sciences under Contract No. DE-AC02-98CH10886. The National Synchrotron Light Source is supported by the U.S. Department of Energy, Division of Materials Sciences and Division of Chemical Sciences. 


\end{document}